\documentclass[aps,prd,showpacs,amsmath,nofootinbib,twocolumn]{revtex4}

\usepackage{graphicx}
\usepackage[latin1]{inputenc}

\newcommand{\be}{\begin{eqnarray}}
\newcommand{\ee}{\end{eqnarray}}
\newcommand{\nn}{~\nonumber \\}

\newcommand{\sfrac}[2]{{\textstyle\frac{#1}{#2}}}
\newcommand{\tr}{\mathrm{tr}}

\usepackage{bbm}

\begin{document}

\title{Constraining vectors and axial-vectors in walking technicolour by a holographic principle}

\author{Dennis D.~Dietrich}

\affiliation{HEP Center, Institute for Physics and Chemistry, University of Southern Denmark, Odense, Denmark}

\author{Chris Kouvaris}

\affiliation{The Niels Bohr Institute, Copenhagen, Denmark}

\date{July 16, 2008}

\begin{abstract}

We use a holographic principle to study the low-energy spectrum of walking
technicolour models. In particular, we predict the masses of the axial
vectors as well as the decay constants of vectors and axial vectors as
functions of the mass of the techni-$\rho$. Given that there are very few
nonperturbative techniques to study strongly coupled theories, using
holography might provide us with insight into how to constrain the
parameters of the low-energy effective action of walking technicolour
models. We also compare our results with findings from other setups.

\end{abstract}


\maketitle

%%%%%%%%%%%%%%%%%%%%%%%%%%%%%%%%%%%%%%%%%%%%%%%%%%%%%%%%%%%%%%%%%%%%%%%%%%

\section{Introduction}

The standard model is remarkably consistent with currently
available data for the interactions of elementary particles, but
it has some theoretical shortcomings. First of all,
although the existence of the (still elusive) fundamental Higgs
scalar is essential for the standard model in order to account for
the spontaneous breaking of the electroweak symmetry,
nature has shown no preference for fundamental scalars in analogous
cases. In the well known examples of superconductivity and
superfluidity, the corresponding scalars turn out to be composite
objects kept together by intricate strong dynamics. Apart from this
aesthetically unappealing feature of the standard model, there
are more serious problems, like the instability of the Higgs
scalar's mass with respect to quantum corrections which
necessitates fine tuning also referred to as the hierarchy
problem. There exists a number of extensions of the standard model
which tries to overcome these deficiencies. The one we are concerned
with here is called technicolour \cite{TC}. In technicolour the
spontaneous breaking of the electroweak symmetry is not due to an
elementary scalar Higgs particle acquiring a vacuum expectation
value. Instead, the standard model without elementary scalar is
supplemented with an additional strongly interacting sector such
that chiral symmetry breaking among the so-called techniquarks
breaks the electroweak symmetry down to the electromagnetic gauge
group. From the experimental point of view, the low-energy
spectrum of such a sector is most accessible and therefore most
relevant. The corresponding degrees of freedom can be encoded in an 
effective Lagrangian which is
constructed such that it reflects the symmetries---gauge and
flavour---of the underlying theory. In general, for a given
 theory, the effective Lagrangian features a large
number of parameters. Linking the effective Lagrangian directly to
the elementary theory is a non-trivial task. In QCD where this
approach is also followed, fitting to experimental data permits to
give values to aforementioned parameters. In the absence of data,
and that, for the time being, is the situation in the case of
technicolour, this is not an option and other principles have to
be devised.

Walking \cite{WTC}, that is quasi-conformal technicolour theories 
with techniquarks in higher dimensional representations are compatible with currently available precision
data \cite{Dietrich:2005jn}. In the following, we will study those walking techicolour models which in
the survey~\cite{Dietrich:2006cm}~have been identified as viable candidates for
breaking the electroweak symmetry dynamically. In particular, we will
concentrate on the prime candidate, usually referred to as minimal walking
technicolour. It consists of two techniflavours transforming under the adjoint
representation of an $SU(2)$ technicolour gauge group. As the adjoint representation
of $SU(2)$ is real, this model features an enhanced flavour
symmetry $SU(4)$
instead of the $SU(N_f)\times SU(N_f)$ for non-(pseudo)real representations. Assuming a
breaking to the minimal diagonal subgroup, the $SU(4)$ breaks to $SO(4)$ which leads to
nine Goldstone modes. This fact together with the enlarged vector and axial vector meson
sector leads to a rich low-energy effective theory \cite{Gudnason:2006ug,Foadi:2007ue}.

There are different ways to constrain the parameter space of the effective Lagrangian.
In \cite{Foadi:2007se}, the authors employ Weinberg's sum rules \cite{Weinberg:1967kj}
for this purpose: Counting the expression for the oblique $S$ parameter
\cite{Peskin:1991sw} as zeroth sum rule, the perturbative
expression obtained for the elementary theory and the value
computed on the low-energy side are postulated to be equal.
Further, the first sum rule is imposed for the respective first
resonances. Likewise, the second, for which also modifications due
to the continuum are taken into account \cite{Appelquist:1998xf}
\footnote{In the strict sense this does not reduce the number of
free parameters, but introduces a new one. Over the latter,
however, one has more quantitative control.}.

Here, we impose a holographic principle---the details
of which will be laid out in the next section---motivated
by the AdS/CFT correspondence \cite{Maldacena:1997re}.
It maps a five-dimensional gravity theory onto a
four-dimensional conformal field theory. Originally, the AdS/CFT
correspondence has been conjectured for $\mathcal{N}=4$ supersymmetry which is an exactly conformal theory. Its application
to non-conformal theories has to be interpreted as
extrapolation. Astonishingly enough, however, holographic descriptions of QCD
give results of remarkably good agreement with experiments~\cite{Erlich:2005qh,Da Rold:2005zs}.
Currently, holography is used extensively as a mathematical tool to probe
aspects of nonperturbative QCD, either in quark-gluon plasma~\cite{Liu:2008tz} or
hadronic physics~\cite{Brodsky:2008pg}. The scale invariance in QCD
is broken due to quantum effects and nonzero quark masses. The success
of the holographic principle in QCD might probably be attributed to the fact that
for energies below 1 GeV, the strong coupling constant effectively behaves as
constant. Lattice simulations support the possibility of an infrared fixed point~
\cite{Furui:2006py,Appelquist:2007hu}. In the case of
technicolour, walking models are by construction quasi-conformal,
where the coupling changes only slightly between the electroweak
scale and the scale of the extended theory which should be larger
than at least 300 times the electroweak. Therefore they meet
the criteria for applying holography in a much better way than QCD
does~\cite{Hong:2006si}. 
Higher-dimensional holographic frameworks have even been used as basis for the construction
of technicolour-like models \cite{Hirn:2006nt}.
One should not take for granted the
findings of the holographic description, but holography remains
one of the very few mathematical tools in the arsenal of
theoretical physics for the study of strongly coupled theories.
Recently, also lattice methods have been implemented for the study
of the dynamics of minimal walking
technicolour~\cite{Catterall:2007yx,DelDebbio:2008wb}. By using
different methods, what one could certainly achieve is a test
of the robustness of investigated features with respect to
different setups.
Within the holographic approach, additional choices must be made,
like, for example for boundary conditions. With the interpretation of the 
fifth dimension as inverse energy scale, there are two very
general types of boundary conditions we impose.
Below, we will compare a hard-wall approach \cite{Erlich:2005qh,PS}, that is the 
fifth dimension $z$ is limited to a finite interval between two values 
$\epsilon$ and $z_m$ that correspond to the ultraviolet and infrared
scales of the theory, respectively, and a soft-wall approach
\cite{Karch:2006pv}, where, although $z$ can go to infinity, there is a potential
term that effectively works as a form factor, smoothing out the sharp boundary 
of the hard-wall model.
For QCD the latter is
phenomenologically favoured as it is able to reproduce the Regge
trajectories for the vector resonances. Due to the absence of the
corresponding data there is no such criterion for the quasi-conformal
theories we are concerned with, and additionally, here we concentrate
on the lowest lying vector and axial-vector resonance, respectively.
Therefore, we compare the two corresponding results on an equal footing.

In Sect.~\ref{II}, we describe the underlying holographic principle, derive
the equations of motion relevant for the treatment of the spectrum of walking technicolour theories,
and give the necessary solutions. In Sect.~\ref{III}, we continue by analysing the
obtained quantitative results and discuss their phenomenological implications.
In Sect.~\ref{IV}, we summarise our findings.

%%%%%%%%%%%%%%%%%%%%%%%%%%%%%%%%%%%%%%%%%%%%%%%%%%%%%%%%%%%%%%%%%%%%%%%%%%

\section{Holographic approach\label{II}}

We will here employ modifications of the five-dimensional holographic models
used in \cite{Erlich:2005qh,Karch:2006pv,Hong:2006si}. In their primary
form, that is for a $SU(N_f)\times SU(N_f)$ flavour symmetry they are
based on the action,
\be
S&:=&\int d^5x \sqrt{g} {\cal L}
\label{action}\\
{\cal L} &:=&
\tr[|D\Phi|^2+m_5^2|\Phi|^2-\sfrac{1}{4g_5^2}(F_L^2+F_R^2)]. \ee
The metric is to be anti de Sitter,
\be ds^2=z^{-2}(-dz^2+dx^2), \ee
with the interpretation of the fifth coordinate, $z$, as inverse
energy scale. The basic idea is to gauge the global group
$SU(N_f)\times SU(N_f)$ in the five dimensional theory. The left,
$A_{L,\mu}^a\hat=\bar q_L\gamma_\mu t^a q_L$, and the right
$A_{R,\mu}^a\hat=\bar q_R \gamma_\mu t^a q_R$, vector fields, 
appear in
the covariant derivative coupling to the scalar $\Phi$,
$D_\mu\Phi:=\partial_\mu\Phi-iA_{L\mu}\Phi+i\Phi A_{R\mu}$, as
well as in the left and right field tensors,
$F_{\mu\nu}:=\partial_\mu A_\nu-\partial_\nu A_\mu
-i[A_\mu,A_\nu]$. $q_{L/R}$ stand for the left-/right-handed
techniquarks. Here, $t^a$ are the generators of the flavour
symmetry group. The scalar field $\Phi^{\alpha\beta}$ corresponds
to the scalar combination of techniquarks $\bar{q}_R^\alpha
q_L^\beta$ \footnote{The case where we have the techniquarks
transforming under a (pseudo)real representation of the gauge
group is slightly different because of an enhanced global symmetry
and is discussed in subsection C.}. By matching to perturbative
calculations the coupling $g_5$ is identified with,
\be g_5^2=12\pi^2/d_\mathrm{R}, \ee
where $d_\mathrm{R}$ is the dimension of the representation of the
technicolour gauge group under which the techniquarks transform.

%-------------------------------------------------------------------------

From the action (\ref{action}), one obtains the equation of motion
for the scalar expectation value $\Phi_0$ which breaks the chiral
symmetry,
\be [z^3\partial_z z^{-3}\partial_z-z^{-2}m_5^2]\Phi_0=0. \ee
With the ansatz $\Phi_0\sim z^d$, one finds the characteristic
equation
\be m_5^2=d(d-4). \label{chareq} \ee
$d$ is equal to the dimension of the scalar operator on the boundary. For a
quasi-conformal theory we set $d=2$, hence, $m_5^2=-4$. The
solution for $\Phi_0$ is
\be \Phi_0=c_1 z^2+c_W z^2\ln(z/\epsilon), \label{confcond} \ee
where $c_1$ and $c_W$ should be fixed by boundary conditions. The
ultraviolet boundary condition is
$(2/\epsilon)\Phi_0(\epsilon)=M$, where $M$ is the techniquark
mass matrix. From this condition, in the chiral limit ($M=0$), we
get $c_1=0$. The other constant $c_W$ will be fixed later on, once
we adjust the technipion decay constant $f_{\pi}$ to the electroweak scale.
%-------------------------------------------------------------------------

The equations of motion for the vector meson, $2V:=A_L+A_R$, is
extracted from the terms of the action (\ref{action}) quadratic in
these fields. In the $V_z=0$ gauge and after having Fourier transformed all 
space-time coordinates except $z$, $V$ satisfies,

\be [z\partial_z z^{-1}\partial_z+q^2]V=0. \ee
%-------------------------------------------------------------------------
Similarly, for the transverse part of the axial vector meson
fields, $2A:=A_L-A_R$,
\be [z\partial_z z^{-1}\partial_z+q^2-g_5^2z^{-2}\Phi_0^2]A=0,
\label{2A} \ee
which couple to the scalar expectation value $\Phi_0$.
For the sake of getting analytic results and with negligible
loss of accuracy, we approximate
$\Phi_0=c_Wz^2\text{ln}(z/\epsilon)\simeq
c_Wz^2\text{ln}(z_m/\epsilon)$. This approximation works well: For
the hard-wall approach which we are going to discuss next, $z$ is
confined (as we already mentioned) between $\epsilon$ and $z_m$.
The approximation we make is worst for small $z$, close to
$\epsilon$, and it gets really good for large $z$, close to $z_m$.
From Eq.~(\ref{2A}), we see that the last term of the
equation is negligible close to the ultraviolet boundary. This means that
the term we approximate can be neglected where the approximation is
not accurate; where it becomes important (close to the infrared
boundary), the approximation is extremely accurate. The same
happens for the soft-wall approach, although, $z$ can in principle assume
arbitrarily large values, the potential term will effectively cut it off 
smoothly. As we shall argue, there is no need to specify a value for $z_m$ 
for the approximation to work because the factor
$\text{ln}(z_m/\epsilon)$ is just a constant and can be absorbed
in $c_W$. $\Phi_0$ from Eq.~(\ref{confcond}) can be written as,
\be \Phi_0\approx Cz^2/g_5. \label{approx} \ee
Thus, in the hard-wall model to be discussed next, $C\approx g_5 c_W
\ln(z_m/\epsilon)$.

%%%%%%%%%%%%%%%%%%%%%%%%%%%%%%%%%%%%%%%%%%%%%%%%%%%%%%%%%%%%%%%%%%%%%%%%%%

\subsection{Hard-wall model}

As mentioned before, in the hard-wall approach $z$ is
confined between $\epsilon$ and $z_m$. Concretely, for the
vector $V$ and axial vector $A$, the wave functions
satisfy the boundary conditions \cite{Erlich:2005qh},
\be
V(\epsilon)=&0&=\partial_z V(z_m),\\
A(\epsilon)=&0&=\partial_z A(z_m). \ee
$z_m$ characterises the position of the infrared boundary,
$\epsilon$ that of the ultraviolet boundary. For
quasi-conformal theories, that is theories which feature almost
conformal behaviour over an interval of scales, the two
aforementioned points along the fifth dimension can be seen as the
boundaries of this interval \cite{Hong:2006si}. For
phenomenologically viable technicolour models $\epsilon\ll z_m$.
To be more precise $z_m/\epsilon$ should be at least 300,
with 1000 being probably a typically expected
value~\cite{Hill:2002ap}. We can, hence, go to the limit
$\epsilon\rightarrow 0$ which turns out to be smooth and not to
have an important quantitative impact.

The pion decay constant $f_\pi$ can be obtained from the solution of the
axial-vector equation of motion for $q^2=0$ and with the boundary
conditions,
\be
\partial_z{\cal A}(z_m)=0\mathrm{~~~and~~~}{\cal A}(0)=1.
\ee
It is given by,
\be g_5^2f_\pi^2=-\partial^2_z{\cal A}(0), \label{fpi} \ee
which arises from $-\epsilon^{-1}\partial_z{\cal A}(\epsilon)$ in the limit 
$\epsilon \rightarrow 0$ because $\partial_z{\cal A}(0)=0$. The general 
solution for the
vectorial equation of motion is given by a linear combination of
the Bessel functions of order one $J_1(qz)$ and $Y_1(qz)$ both
multiplied by $z$. Only $zJ_1(qz)$ satisfies the boundary
condition at $z=0$. The boundary condition at $z=z_m$ implies,
\be J_0(qz_m)=0, \ee
which represents an eigenvalue equation for $q$. Thus, the mass of
the lightest vector resonance, the techni-$\rho$, is given by,
\be M_V=2.4048/z_m, \ee
where the numerator is given by the first zero of $J_0$.

The axial equation of motion is solved by linear combinations of
$z^2e^{-Cz^2/2}$ times the Kummer functions (confluent
hypergeometric functions) $M(1-\sfrac{q^2}{4C},2,Cz^2)$ and
$U(1-\sfrac{q^2}{4C},2,Cz^2)$. The boundary condition at $z=0$ eliminates
contributions involving the function $U$. Then the boundary condition at
$z=z_m$ leads to the eigenvalue equation,
\be (2C^2z_m^2-M_A^2)M(1-\sfrac{M_A^2}{4C},2,Cz_m^2) &+& \nn +
(4C+M_A^2)M(-\sfrac{M_A^2}{4C},2,Cz_m^2) &=& 0, \ee
which can only be evaluated numerically.

The differential equation for ${\cal A}$ possesses solutions made
up of hyperbolic functions, $\sinh(Cz^2/2)$ and $\cosh(Cz^2/2)$.
The boundary condition at $z=0$ fixes the prefactor for the
$\cosh$ term to 1. Exploiting the expression arising from the
boundary condition at $z=z_m$ leads to,
\be {\cal A} = \cosh(Cz^2/2) - \tanh(Cz_m^2/2)\sinh(Cz^2/2).
\label{acal} \ee
Evaluation of Eq.~(\ref{fpi}) for this solution yields,
\be g_5^2f_\pi^2=C\tanh(Cz_m^2/2). \ee

The decay constants of the vector and the axial vector are obtained from
\be
g_5F_V&=&\partial_z^2 V(0),\\
g_5F_A&=&\partial_z^2 A(0), \ee
where $V$ and $A$ have to be normalised according to,
\be 
\int_0^{z_m}\frac{dz}{z}V^2 = 1 = \int_0^{z_m}\frac{dz}{z}A^2 .
\ee
The normalised vectorial solution reads,
\be V=\sqrt{2}\frac{zJ_1(M_Vz)}{z_mJ_1(M_Vz_m)}, \ee
and leads to the decay constant,
\be g_5F_V=1.1328~M_V^2, \label{fvhard} \ee
where the expression has been evaluated at the first zero of $J_0$. The
normalisation integral for $A$ must be evaluated numerically.

%%%%%%%%%%%%%%%%%%%%%%%%%%%%%%%%%%%%%%%%%%%%%%%%%%%%%%%%%%%%%%%%%%%%%%%%%%

\subsection{Soft-wall model}

In \cite{Karch:2006pv} an additional dilaton field $\phi$ is introduced into the action,
\be S_s := \int d^5x\sqrt{g}e^{-\phi}{\cal L}. \ee
Requiring that the mass spectrum show a Regge like spacing
linear in the squared mass, the dilaton background should behave
like $cz^2$ ($c=\mathrm{constant}$) for large values of $z$. Compared to
the potential well of the hard-wall model, this leads to a
harmonic oscillator like setting which in turn gives linearly
spaced eigenvalues for the squared mass.

Rederiving the vectorial equation of motion yields,
\be [ze^{+cz^2}\partial_z z^{-1}e^{-cz^2}\partial_z+q^2]V_s=0, \ee
and the infrared boundary condition is replaced by postulating the normalisability of the solution over $\mathbbm{R}^+$.
With the substitution,
\be V_s=:v_se^{+cz^2/2}, \ee
$v_s$ obeys the equation of motion,
\be [z\partial_z z^{-1}\partial_z+q^2-c^2z^2]v_s=0, \ee
and must satisfy the same boundary condition as $V_s$. The
differential equation for $v_s$ coincides with that for $A$ in
the hard-wall case, up to the interchange of the constants $c$
and $C$. Therefore, the solution for $v_s$ with the correct
behaviour at the UV boundary has already been given in the
previous section in the context of the axial vector wave function.
The aforementioned normalisability
 of the solution implies that the eigenvalues $M_V^2$ be integer multiples of $4c$.
 (For these values the Kummer function $M$ turns into a polynomial.)

Before we can continue with the axial vector mesons, we have to know the
expectation value $\Phi_s$ for the scalar. The relevant equation of motion is given
by,
\be [z^3e^{+cz^2}\partial_z
z^{-3}e^{-cz^2}\partial_z+z^{-2}m_5^2]\Phi_s=0. \ee
With a substitution,
\be \Phi_s=:\varphi_s e^{+cz^2/2}, \label{xyz} \ee
which leads to the same boundary condition for $\varphi_s$ as for $\Phi_s$, the previous differential equation turns into,
\be [z^3\partial_z
z^{-3}\partial_z+z^{-2}m_5^2-c^2z^2]\varphi_s=0. \ee
For $c^2z^4\ll m_5^2$ the characteristic equation (\ref{chareq})
holds to good approximation and through identification in the ultraviolet we can again set $m_5^2=-4$. The solution for the previous equation is then given by linear
combinations of $z^2$ times the Bessel functions $I_0(cz^2/2)$ and
$K_0(cz^2/2)$. (For $m_5^2\neq -4$ the order of the Bessel
functions changes.) 
The boundary condition $\lim_{\epsilon\rightarrow 0}\Phi_s(\epsilon)=0$ and equivalently $\lim_{\epsilon\rightarrow 0}\varphi_s(\epsilon)=0$ selects the solution,
\be
\varphi_s=c_sz^2I_0(cz^2/2) .
\ee
Already $\varphi_s$ exhibits exponential growth for large values of $z$ and moreso does $\Phi_s$ with its additional exponential factor. This observation is indicative of an instability which must eventually be intercepted by non-linear terms in the equation of motion which would originate from potential terms involving the (pseudo)scalar fields \cite{Karch:2006pv}. At the linear level we have to adhere to small values of $z$ and we can use henceforth,
\be
\varphi_s=c_sz^2[1+O(c^2z^4/4)] .
\ee
$\Phi_s$ contains an additional exponential factor. As we are confined to small values of $z$ in any case and for the sake of an analytical result we approximate it by unity,
\be
\Phi_s=c_sz^2[1+O(cz^2/2)].
\label{approximation}
\ee
This is the dominant approximation and we shall determine its range of applicability below.

The equation of motion for the wave function of the axial vector mesons reads,
\be [ze^{+cz^2}\partial_z z^{-1}e^{-cz^2}\partial_z +
q^2-g_5^2z^{-2}\Phi_s^2]A_s=0, \ee
where $\Phi_s$ is the expectation value for the scalar. As above, we carry out the substitution,
\be A_s=a_se^{+cz^2/2}, \label{subas} \ee
which leads to,
\be [z\partial_z z^{-1}\partial_z +
q^2-c^2z^2-g_5^2z^{-2}\Phi_s^2]a_s=0, \ee
and does not affect the boundary conditions.

Using the dominant term from Eq.~(\ref{approximation}), we end up with the 
following equation of motion for $a_s$,
\be \{z\partial_z z^{-1}\partial_z +
[q^2-(c^2+C^2)z^2]\}a_s=0. \ee
Up to the replacement of $c^2$ by $c^2+C^2$, it coincides with the equation
of motion for $v_s$. The boundary condition is the same. Thus, the 
corresponding solution can be obtained by carrying out the aforesaid replacement. 
Accordingly, requiring the normalisability of the wave function implies that
the squared mass eigenvalues $M^2_A$ be integer multiples of $4\sqrt{c^2+C^2}$.

Finally, we would like to extract the pion decay constant from the solution
of the axial differential equation with $q^2=0$,
\be (ze^{+cz^2}\partial_z z^{-1}e^{-cz^2}\partial_z-C^2z^2){\cal
A}_s=0, \ee
or
\be [z\partial_z z^{-1}\partial_z-(c^2+C^2)z^2]\mathbbm{A}_s=0,
\ee
where,
\be {\cal A}_s=:\mathbbm{A}_se^{+cz^2/2}. \ee

Replacing $C$ by $\sqrt{C^2+c^2}$ in Eq.~(\ref{acal}) and sending $z_m$ to
infinity yields,
\be \mathbbm{A}_s=e^{-z^2\sqrt{C^2+c^2}/2}, \ee
or
\be {\cal A}_s=e^{-z^2(\sqrt{C^2+c^2}-c)/2}. \ee
Hence, from Eq.~(\ref{fpi}),
\be g_5^2f_\pi^2=\sqrt{C^2+c^2}-c. \label{fpi2} \ee
Using the previous results on the mass eigenvalues we find,
\be M_A^2=M_V^2+4g_5^2f_\pi^2. \label{hyperbola} \ee
(For the $n^\mathrm{th}$ pair of resonances this relation turns into
$M_{A,n}^2=M_{V,n}^2+4ng_5^2f_\pi^2$.)

In order to extract the decay constants for the vector and axial vector
mesons it remains to normalise the corresponding wave functions according
to,
\be
\int_0^\infty\frac{dz}{z}v_s^2(z)=1=\int_0^\infty\frac{dz}{z}a_s^2(z) . 
\ee
With $M_V^2=4c$ the normalised wave function for the vector reads,
\be V_s=\sqrt{2}cz^2. \ee
The decay constant equals,
\be g_5F_V=\partial_z^2 V_s(0)=2\sqrt{2}c=M_V^2/\sqrt{2}.
\label{fvsoft} \ee
With a prefactor of $0.707\dots$ the rise of $F_V$ with $M_V^2$ is shallower
than in the hard-wall model, where the prefactor was approximately $1.133$.
Similarly, the results for the axial vector are,
\be A_s=\sqrt{2}\sqrt{c^2+C^2}z^2, \ee
and
\be g_5F_A=\partial_z^2
A_s(0)=2\sqrt{2}\sqrt{c^2+C^2}=M_A^2/\sqrt{2}. \label{fasoft} \ee

%=========================================================================

As already announced, we are now going to assess the range of applicability
of the approximations carried out above. The approximation introducing an
$O(cz^2)$ deviation was replacing the additional exponential in
Eq.~(\ref{xyz}) by unity. In the spirit of perturbation theory, we now
calculate the shift $\delta q^2$ of $M_A^2$ induced by the perturbation,
that is the difference between the exact and the approximated potential. For
this we need to know the normalised wave function,

\be
a_s=\sqrt{2}\sqrt{c^2+C^2}z^2e^{-\sqrt{c^2+C^2}z^2/2},
\label{wf}
\ee
and the perturbation of the potential,
\be
\delta U := C^2z^2(e^{+cz^2}-1).
\ee 
The shift is then given by,
\be
\delta q^2
&=&
\int_0^\infty\frac{dz}{z}~a^2_s~\delta U
=
\nn
&=&
2C^2(C^2+c^2)
\times
\nn
&&\times
[(\sqrt{C^2+c^2}-c)^{-3}-(C^2+c^2)^{-3/2}].
\ee 
This quantity must be compared to $M_A^2=4\sqrt{c^2+C^2}$ and should be
smaller than the latter, implying that,
\be
x^2\sqrt{1+x^2}[(\sqrt{1+x^2}-1)^{-3}-(1+x^2)^{-3/2}]\ll 2 ,
\nn
\ee
where $x:=C/c$. The left-hand side of the previous expression diverges for
small values of $x$ and tends to zero for large values of $x$. Therefore, we
must have $C\gg c$. With the help of Eq.~(\ref{fpi2}) we can translate this
into ranges for the vectorial mass $M_V$,

\be
x=\sqrt{(1+y^2)-y^2}/y ,
\ee
where $y:=c/(g_5^2f_\pi^2)$. The previous expression tends to infinity for small values of $y$ and to zero for large values of $y$. We should, thus, keep $y$ small. This corresponds to an upper bound on the vectorial mass, $M_V=2\sqrt{c}\ll g_5f_\pi$. For minimal walking technicolour this corresponds to roughly 1.5TeV. Without giving the details here, in the spirit of an iterative procedure,
we have further improved the treatment of the soft-wall setting by calculating a new effective value for the parameter $C$ by averaging the corresponding potential term with the initially obtained wave-function (\ref{wf}). Comparison of the initial with the thus obtained result shows that at very small $M_A$ the two results coincide and that at $M_A\approx1.5$TeV the old result is ten percent lower than the new. For $M_A\approx 3$TeV the results differ by a hundred percent. 

%%%%%%%%%%%%%%%%%%%%%%%%%%%%%%%%%%%%%%%%%%%%%%%%%%%%%%%%%%%%%%%%%%%%%%%%%%

\subsection{Real and pseudo-real representations}

For $N_f$ techniquarks transforming under real or pseudo-real
representations of the technicolour gauge group, the unbroken
flavour symmetry is enhanced to $SU(2N_f)$. The
$SU(N_f)_L\times SU(N_f)_R$ which has been treated up to this
point is a subgroup of $SU(2N_f)$ and can be embedded in it.
There will not only be the previous $2(N_f^2-1)$ (axial) vectors,
but a total of $(2N_f)^2-1$, that is an extra $2N_f^2+1$. In the
action they will have their own kinetic terms and they appear in
the covariant derivative. Apart from new characteristics like
non-zero technibaryon number, from the point of view of the
flavour symmetry they are still either vector or axial vector
eigenstates. Hence, if the unbroken flavour symmetry is an exact
symmetry, they have the same masses as the standard vectors and 
axial vectors, respectively. If they
couple to the condensate (axial eigenstate) or not (vectorial
eigenstate) depends on whether the corresponding generator of the
flavour symmetry group commutes with the condensate or not.
Such a coincidence of values is also present for the decay constants of the states which can decay, that is the mesons. At the present level, technibaryon number is conserved, so that the baryonic modes cannot decay.

In order to give a concrete example, let us look at minimal
walking technicolour, using the notation from \cite{Foadi:2007ue}.
There are two techniquarks, $U$ and $D$, that transform under
the adjoint representation of the $SU(2)$ technicolour group.
As the adjoint of $SU(2)$ is a real representation, the
global symmetry of the model is $SU(4)$. The flavour symmetry no
longer only transforms left-fields, $(U_L,D_L)$, among themselves
and right-fields, $(U_R,D_R)$, among themselves, but also left-
into right-fields and vice versa. Therefore, it is practical to
work in a joint basis, say
\be
Q_j^{\alpha}=
\left(
\begin{array}{c}
U_{L} \\
D_{L} \\
-i\sigma^2 U_{R}^* \\
-i\sigma^2 D_{R}^*
\end{array}
\right),
\ee
where Greek and Latin indices denote spin ($\alpha$ runs from 1 to 2) and flavour components,
respectively. Let $T^a$ be a set of 15 generators for $SU(4)$ in the fundamental representation. Divide them into the first six, $S^a:=T^a$, $a\in\{1;\dots;6\}$, commuting with the condensate, which is to be proportional to a $4\times 4$ matrix $E$, and the following nine, $X^a:=T^{a+6}$, $a\in\{1;\dots;9\}$, which do not commute with $E$. For
\be
E=\left(
\begin{array}{cc}
0 & \mathbbm{1} \\
\mathbbm{1} & 0
\end{array}
\right),
\ee
where $\mathbbm{1}$ stands for the $2\times 2$ unit matrix, the condensate is thus,
$\langle Q_i^\alpha Q_j^\beta\epsilon_{\alpha\beta}E^{ij}\rangle
=
-2\langle\bar{U}_R U_L+\bar{D}_R D_L\rangle$, where $\epsilon_{\alpha\beta}$ is the two-dimensional antisymmetric symbol.
(For an explicit realisation for the generators see Appendix A of
\cite{Foadi:2007ue}.) The covariant derivative for $\Phi$ in
the action~(\ref{action}) is given by
\be D_{\mu} \Phi =
\partial_{\mu} \Phi - iA_{\mu} \Phi -i \Phi A_{\mu}^{\top}, \ee
where $A_{\mu}=A^a_\mu T^a$. $\Phi^{\alpha \beta}$ is also promoted to a $4 \times 4$ matrix,
that now represents the techniquark combination
$Q_i^{\alpha}Q_j^{\beta}\epsilon_{\alpha\beta}$. It should be noted that the
above techniquark composite scalar contains the degrees of freedom associated with $\bar{q}_R^\alpha q_L^\beta$ known from the previous sections, exactly as $SU(4)$
contains $SU(2)_L \times SU(2)_R$ as subgroup. The
$SU(2)_L\times SU(2)_R$ part can be embedded by defining the
corresponding generators according to $\sqrt{2}L^a:=X^a+S^a$ and
$\sqrt{2}R^a:=X^a-S^a$, where $a\in\{1;2;3\}$. The fields belonging to the
generators $S^4$, $S^5$, and $S^6$ lead to three additional
eigenstates. In this special setting one of them is a meson and the other
two an axial vector baryon and its antibaryon. The fields
belonging to the six remaining generators $X^4$ to
$X^9$ not commuting with the condensate correspond to three baryon-antibaryon 
pairs which are vector
eigenstates. In addition to the three mesonic Goldstone modes
arising from the broken axial $SU(2)$, there are six more,
arranged in three baryon-antibaryon pairs. They have also to be
included in the corresponding kinetic and mass terms. We
assume zero mass for the techniquarks, and, therefore, there is no
isospin splitting. Hence, the richer pseudoscalar sector does not
influence the results at this level. What the decay constants of
the technibaryonic states---spin zero as well as spin-one---are
concerned, these states do not decay, as on this level
technibaryon number is conserved \footnote{That is, conserved
perturbatively. It can change through sphaleron processes which,
however, are extremely suppressed \cite{Gudnason:2006yj}.}.

%%%%%%%%%%%%%%%%%%%%%%%%%%%%%%%%%%%%%%%%%%%%%%%%%%%%%%%%%%%%%%%%%%%%%%%%%%

\section{Results and discussion\label{III}}

\subsection{Minimal walking technicolour}

After fixing $f_\pi$ to its phenomenological value of
$246\sqrt{2/N_f}$ GeV, there is only one free parameter left in
the present approach. Consequently, the results can be presented
as one-parameter curves. For minimal walking technicolour with its
$N_f=2$ flavours in the (three-dimensional) adjoint representation
of $SU(2)$, $g_5^2=4\pi^2$.

\begin{figure}[t]
\includegraphics[scale=0.81]{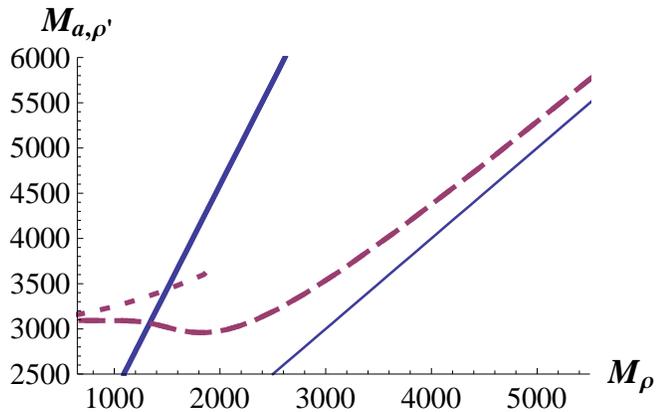}
\caption{Minimal walking technicolour: The mass $M_a$ of the
first axial vector meson as function of the mass $M_\rho$ of the
first vector meson in the hard-wall model (long dashes) and in the
soft-wall model (short dashes). For comparison, the finer
straight line indicates the diagonal $M_a=M_\rho$. The wider
straight line indicates the mass $M_{\rho^\prime}$ of the second
vector resonance in the hard-wall model. } \label{Fig1}
\end{figure}

The axial vector meson mass as a function of the vector meson mass
is shown in Fig.~\ref{Fig1}. For small vector meson masses, the
mass of the first axial vector state coincides in the two
scenarios, hard- and soft-wall, respectively. The axial vector
mass is bounded from below. The axial vector mass in the soft-wall
model is a monotonously rising function of the vector mass and,
hence, the minimum is reached for vanishing vector masses. It is
given by,
\be M_a>2g_5f_\pi, \ee
which here equals roughly 3 TeV. (From this point onward, we
will use the subscript $a$ for the lightest axial vector resonance
and $\rho$ for the lightest vector resonance. Likewise,
${\rho^\prime}$ indicates the corresponding first excitation.) In
the hard-wall model, $M_a$ starts out by first decreasing slightly
when $M_\rho$ is increased before increasing again. In both
scenarios the axial vector mass then approaches the vector mass
from above without ever falling below it. The axial vector mass in 
the soft-wall approach stays also
always larger than the hard-wall approach. For a light first
vector meson ($M_\rho$) also the second resonance
($M_{\rho^\prime}$)---or even higher vector resonances---can be
lighter than the first axial vector meson (see the bold straight
line in Fig.~\ref{Fig1}).

%=========================================================================
%=========================================================================

\begin{figure}[t]
\includegraphics[scale=0.81]{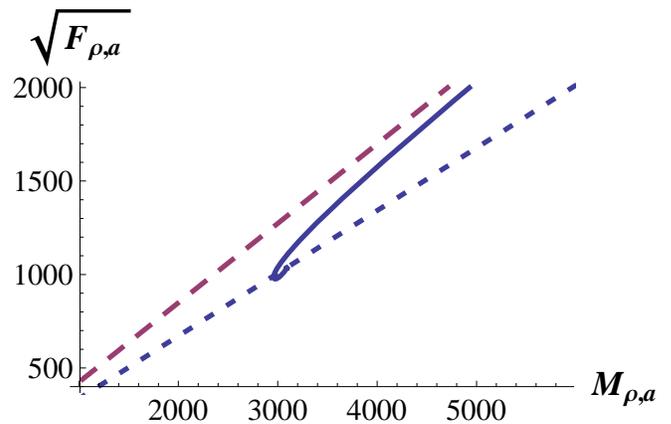}
\caption{Minimal walking technicolour: The square-root of the decay
constant of the first vector meson $\sqrt{F_\rho}$ as function of
its mass $M_\rho$ (hard-wall: long dashes, soft-wall: short dashes) and the
square-root of the decay constant of the first axial vector meson
$\sqrt{F_a}$ as function of its mass $M_a$ (hard-wall: solid line,
soft-wall: short dashes) for the hard- and the soft-wall model,
respectively. The hook-like structure in the axial vector graph is
linked to the non-monotonous behaviour of $M_a$ as a function of
$M_{\rho}$ in the hard-wall model (see Fig.~\ref{Fig1}). In the soft-wall model the graphs for the vector and axial vector coincide. Remember, however, that there the approximations made for the {\it axial} vector are only valid for small masses.
} \label{Fig2}
\end{figure}

In the soft-wall model, the decay constants both for the vector
and the axial vector meson as functions of the respective mass
show the same behaviour, that is linear for the square root of the
decay constants (see Fig.~\ref{Fig2}). The same is true for the
vector meson in the hard-wall model, albeit with a different
slope. For the hard-wall model the decay
constant of the axial vector shows a different behaviour. For increasing mass, it
approaches the vectorial behaviour from below.
At small values of the mass the axial vector decay constant in the
hard-wall model is numerically closer to the outcome of the
soft-wall scenario. This is understandable as for large masses
in the hard-wall scenario the spacing between the infrared and the
ultraviolet wall is rather small and the tilting of the floor of
the square-well potential by the condensate for the axial vector
as opposed to the flat bottom for the vector does not play a
prominent role. For small masses the infrared wall is
approximately infinitely far away and the axial behaves like in
the soft-wall case. This is also the reason why its mass does not
go down to arbitrarily small values in either scenario.

%=========================================================================
%=========================================================================

The setup laid out in Ref.~\cite{Foadi:2007ue}, contains
additional terms coupling scalar operators to the spin-one fields
not present in the holographic approach pursued here.
\{See, for example, Eq.~(41) in Ref.~\cite{Foadi:2007ue}.\} 
This is chiefly due to the fact that there the spin-one fields are associated with the {\it global} flavour symmetry. This allows for more invariant terms than if the field is associated with a {\it local} symmetry as is the case in a holographic setting. A non-minimal term, which is also locally invariant is $\tr(F_{\mu\nu}\Phi F^{\mu\nu\top}\Phi^\dagger)$ \cite{Sannino:2008ha}. 
The higher number of parameters allows for a more diverse
phenomenology than seen in the present, more (and differently)
constrained approach. Most prominently, the axial vector meson may
also be lighter than the vector meson, in contrast to the present
findings. Here, the axial is found to have a mass of $\sim 3$ TeV
and above. In this range also the mass hierarchy in
Ref.~\cite{Foadi:2007se} is the one known from quantum
chromodynamics with the lighter vector.
Our Eq.~(\ref{hyperbola}) is directly reminiscent of Eq.~(C8) in Ref.~\cite{Foadi:2007ue}.
This allows us to compare our $\sqrt{C^2+c^2}-c$ to their $v^2\tilde g^2 r_2/8$.
There $v^2$ stands for the strength of the chiral condensate, $\tilde g$ represents
the coupling constant for the (axial) vector fields, and $r_2$ parametrises
the relative strength of one particular contribution term coupling the (axial)
vectors to the (pseudo) scalars. \{See Eq.~(41) in \cite{Foadi:2007ue}.\}
In our setup there is a fixed link, Eq.~(\ref{hyperbola}), between the masses
$M_a$ and $M_\rho$ and the pion decay constant $f_\pi$. This is the reason why here fixing
$f_\pi$ to its physical value constrains the axial vector mass from below. In
Ref.~\cite{Foadi:2007ue} another term (with relative strength $r_3$)
influences $f_\pi$ and the direct impact of fixing $f_\pi$ to its physical value on the 
axial mass is softened.

%%%%%%%%%%%%%%%%%%%%%%%%%%%%%%%%%%%%%%%%%%%%%%%%%%%%%%%%%%%%%%%%%%%%%%%%%%

\subsection{Beyond minimal walking technicolour}

%-------------------------------------------------------------------------
\begin{table}[t]
\begin{tabular}{ccccccc}
$N_c$&representation&$d_\mathrm{R}$&$N_f$&$N_f^g$&$M_{a,\mathrm{soft}}^\mathrm{min}$&$(F_{a}^\mathrm{min})^{1/2}$\\
\hline
2&fundamental&2&7&6&2.2&0.66\\
2&fundamental&2&7&2&3.8&1.15\\
2&adjoint&3&2&2&3.1&1.04\\
3&fundamental&3&11&2&3.1&1.04\\
3&2-ind.sym.&6&2&2&2.2&0.87\\
3&adjoint&8&2&2&1.9&0.81\\
4&fundamental&4&15&2&2.7&0.97\\
4&2-ind.sym.&10&2&2&1.7&0.77\\
4&2-ind.antisym.&6&8&2&2.2&0.87\\
4&adjoint&15&2&2&1.4&0.69\\
5&fundamental&5&19&2&2.4&0.91\\
5&2-ind.antisym.&10&6&2&1.7&0.77\\
6&fundamental&6&23&2&2.2&0.87\\
\hline
&&&&&TeV&TeV
\end{tabular}
\caption{
Various walking technicolour models from Tab.~III in \cite{Dietrich:2006cm}.
Minimal axial vector meson mass $M_{a,\mathrm{soft}}^\mathrm{min}$ and square-root of
the minimal axial decay constant $(F_{a}^\mathrm{min})^{1/2}$
for various walking technicolour models characterised by the representation
of the technicolour gauge group under which the techniquarks transform and
the number of (gauged) techniflavours $N_f$ ($N_f^g$). 
}
\label{Tab1}
\end{table}
%-------------------------------------------------------------------------

\begin{figure}[t!]
\includegraphics[scale=0.81]{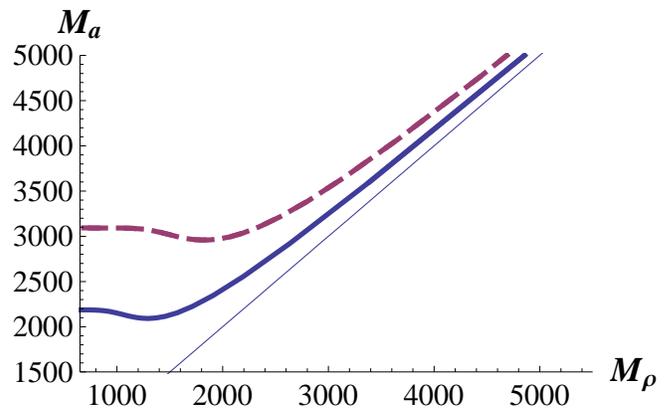}
\caption{
$M_a$ as function of $M_\rho$ in the hard-wall model for minimal walking
technicolour (dashed) and for the model with seven techniquarks (six of them gauged 
under the electroweak interactions)
in the fundamental representation of $SU(2)$ (solid). The straight line indicates 
the value of $M_\rho$ for comparison.
}
\label{Fig3}
\end{figure}

In Ref.~\cite{Dietrich:2006cm} other models beyond minimal walking
technicolour were listed systematically. These are viable
candidates for dynamical electroweak symmetry breaking, as they
may display a sufficiently large amount of walking and are not at
odds with electroweak precision data. In view of the present
computation they are characterised by the number of techniflavours
$N_f$ and the dimension $d_R$ of the representation $R$ of the
gauge group under which the techniquarks transform. Conveniently,
different values of these parameters lead only to rescalings of
the axes of the plots in Fig.~\ref{Fig1}. Concretely, the masses
are multiplied by $(3/d_R)^{1/2}$ and $(2/N_f)^{1/2}$. This scales
the $M_a$ graph downwards and the dip visible in the hard-wall
model to the left when we increase $d_R$ and/or $N_f$ (see Fig.~\ref{Fig3}). In this context it is important to point out that $N_f$ {\it in this scaling} is given by the number of techniflavours gauged under the electroweak group. First of all, this must be an even number to have complete doublets. Then it has proven to be advantageous to include more than two techniquarks to be close to conformality, while only gauging two under the electroweak gauge group to alleviate the bounds from electroweak precision data. 
For the {\it partially gauged} \cite{Dietrich:2005jn,Dietrich:2006cm} 
technicolour models, the number of gauged flavours is indicated in Tab.~\ref{Tab1} in the column marked by $N_f^g$.
As it turns out, of all models, the minimal mass for the axial vector meson in minimal walking 
technicolour is only surpassed in the partially gauged
(two flavours gauged, seven overall) model with techniquarks in the 
fundamental representation of $SU(2)$. 
For all its sibling models the mass is reduced (see Tab.~\ref{Tab1}).

%=========================================================================
%=========================================================================

The decay constants of the spin-one mesons scale differently, courtesy of
the extra factor of $g_5$ in Eqs.~(\ref{fvsoft}) and (\ref{fasoft}). In both
models and for vectors as well as axial vectors, the $F^{1/2}$ scale like
$(3/d_R)^{1/4}$ and $(2/N_f)^{1/2}$. As measure for this scaling the
rightmost column in Tab.~\ref{Tab1} shows the square-root of the minimal
value for the decay constant of the axial vector. It is achieved in the
soft- as well as the hard-wall model when $M_\rho\rightarrow 0$.

Of all the models listed in Tab.~\ref{Tab1} only those with
techniquarks in the two-index symmetric representation of $SU(3)$
and $SU(4)$, respectively, feature the most basic flavour symmetry
breaking pattern, that is $SU(2)_L\times SU(2)_R\rightarrow
SU(2)_V$, which gives rise only to the (mesonic) fields contained
in the action (\ref{action}) and treated in detail in what
followed.

All models with techniquarks in the adjoint representation possess two
flavours irrespective of the number of colours. Thus, they are all covered
exactly by the above discussion up to the rescaling of the mass eigenvalues
and decay constants treated in the present subsection.

The models with techniquarks in the fundamental representation of $SU(3)$ to
$SU(6)$ and in the two-index symmetric representation of $SU(5)$ have an
enlarged flavour symmetry due to their number of flavours $N_f$ being larger
than two, but not because of a (pseudo)real
representation. Hence, we encounter $SU(N_f)_L\times SU(N_f)_R\rightarrow
SU(N_f)_V$. The symmetry does not mix left and right fields. Consequently,
we have only mesonic states among the (pseudo)scalars and (axial) vectors.
The decomposition into vector and axial vector eigenstates is 
straightforward, $2V=A_L+A_R$ and $2A=A_L-A_R$.

The remaining model, seven flavours of the fundamental representation of
$SU(2)$ as well as eight flavours of the two-index antisymmetric
representation of $SU(4)$ feature enhanced symmetries due to the fact that
they contain more than two flavours, but also because they feature pseudoreal
and real representations, respectively. Hence, once more, part of the flavour
symmetries involve mixing left and right handed fields and part of the extra
spin-zero and spin-one states carry technibaryon number. Nevertheless, the spin-one 
particles are always either vector or axial vector eigenstates when
described within the present framework. The exact decomposition
involves handling $SU(14)$ or $SU(16)$ generators and is not performed here.

%%%%%%%%%%%%%%%%%%%%%%%%%%%%%%%%%%%%%%%%%%%%%%%%%%%%%%%%%%%%%%%%%%%%%%%%%%

\section{Summary\label{IV}}

We have constrained the parameter space for the effective
low-energy description for a set of walking technicolour models by
imposing different variants of a holographic principle. In
general, in such an effective action approach to strongly
interacting theories, it is rather difficult to link the
abundantly arising parameters to the less numerous parameters in
the elementary theory. In the case of QCD, one can resort to
experimental data. In beyond the standard model physics, this
input is to date at best sparse. Here, one can fall back on a set
of postulates which further constrain the model and can serve to
enhance its predictive power. We chose two variants of a
holographic principle. The holographic approaches have been
adapted from Ref.~\cite{Erlich:2005qh}~for the hard-wall model and
Ref.~\cite{Karch:2006pv}~for the soft-wall model. The soft-wall
model was introduced in order to model the equally spaced
mass-squared eigenvalues known approximately from QCD (Regge
trajectories) and expected in the presence of linear confinement.
As it is not clear what to expect at the low-energy end for a
quasi-conformal theory, especially for those with matter in the
adjoint representation of the gauge group, we here simply compare
the results from the hard- and the soft-wall scenario. There is
accordance of these holographic descriptions with experimental
values found for QCD, although QCD is a running theory and the
holographic principle is based on a conjecture for a conformal
theory. Clearly, there exists no rigorous derivation for the here
used low-energy description from the elementary theory. However,
in view of the match achieved for QCD, a better accordance may be
expected for the almost conformal technicolour theories we are
concerned with. Seeing the potentially achievable
predictive power, efforts towards the better understanding of the
correspondence would appear to be a good investment.

The technicolour models were taken from the list of models
which in Ref.~\cite{Dietrich:2006cm} have been identified as viable candidates
for breaking the electroweak symmetry dynamically, passing currently available
electroweak precision tests. Among these is also the prime candidate,
minimal walking technicolour, with two techniquarks transforming under the
adjoint representation of $SU(2)$.

In the present approach the results obtained for mass
eigenvalues and decay constants can be linked by scaling laws linking the results between different theories and we can
concentrate on a single one in this synopsis. The masses scale like $d_\mathrm{R}^{-1/2}$
with the representation of the technicolour group with respect to which the
techniquarks transform and like $N_f^{-1/2}$ with the number of flavours.
The square-roots of the decay constants, which in the conventions used here have the
dimension of mass, scale like $d_\mathrm{R}^{-1/4}$ and again like
$N_f^{-1/2}$.
For small vectorial masses the axial vector mass is bounded from below; in
minimal walking technicolour at about 3 TeV and changed according to the
scaling laws for the others (see Tab.~\ref{Tab1} and Fig.~\ref{Fig3}). For
large values of the
vectorial mass also the mass of the lightest axial vector rises, becomes
ever closer to the vector mass, but never falls below it (see Fig.~\ref{Fig1}).
For a light first vector resonance also the second or even more vectors may
be lighter than the first axial vector.

The predictions for the decay constants also differ somewhat
between the two setups. In the soft-wall model the square-roots of
the decay constants are linear functions of the corresponding mass
with the same slope and zero intercept. In the hard-wall model for
the vector, the relation is also linear with zero intercept, but
larger slope. There, for the lowest possible values of the mass,
the axial vector decay constant starts out with a value close to
the soft-wall model and approaches the vector result in the
hard-wall model from below for large values of the mass.

In other descriptions based on less and/or differently constrained
effective low-energy actions these features can be different. For
example for the setup in Refs.~\cite{Foadi:2007ue, Foadi:2007se}
also inverted mass hierarchies can be found, with the first axial
vector lighter than the first vector. In the range where the axial
vector is as heavy as 3 TeV or heavier, the mass hierarchy is,
however, predicted to be the normal one. In our approach fixing
the pion decay constant to its physically required value
automatically puts a lower bound on the axial vector mass. In the
other approach, this is buffered by the larger number of
parameters. As outlook, it would be interesting to investigate the
role the corresponding terms play when combined with a holographic
recipe. Further, the inclusion of non-zero quark masses may prove
insightful.

%%%%%%%%%%%%%%%%%%%%%%%%%%%%%%%%%%%%%%%%%%%%%%%%%%%%%%%%%%%%%%%%%%%%%%%%%%

\section*{Acknowledgments}

The authors would like to thank Roshan Foadi, Mads T.~Frandsen, Deog-Ki Hong,
Matti J\"arvinen, and Francesco Sannino for discussions. The work
of DDD was supported by the Danish Natural Science Research
Council. The work of CK was supported by the Marie Curie
Fellowship under contract MEIF-CT-2006-039211.

%%%%%%%%%%%%%%%%%%%%%%%%%%%%%%%%%%%%%%%%%%%%%%%%%%%%%%%%%%%%%%%%%%%%%%%%%%

\end{document}